# Ultra-high field enhancing in Split Ring Resonators by azimuthally polarized excitation

**Jacob Scheuer**

*School of Electrical Engineering Tel Aviv University, Ramat Aviv, Tel-Aviv 69978, Israel*
*kobys@eng.tau.ac.il*
*www.eng.tau.ac.il/~kobys/*

**Abstract:** We study the field enhancement and resonance frequencies in split-ring resonators (SRR) illuminated by azimuthally polarized light. We find that compared to linearly polarized illumination, the azimuthally polarized illumination increase the intensity enhancement by more than an order of magnitude. We attribute the increase in the intensity enhancement to the improved overlap between the SRR geometry and the direction of the electric field vector in each point. In addition, we present and explore a method to tune the resonance frequency of the SRR (for azimuthal polarization) by introducing more gaps to the structure. This approach allows for simple and straightforward tuning of the resonance frequency with small impact on the intensity enhancement. The impact of the design parameters on the intensity enhancement under azimuthally polarized illumination is also studied in details, exhibiting clear differences to the case of linear polarized illumination.

## 1. Introduction

The ability to focus and concentrate high optical intensities in ultra-small, sub-diffraction limits, volumes using plasmonic devices and effects has drowned much attention during the last decade [1-22]. In addition to its scientific importance, nanometer scale focusing is an enabling technology for numerous applications such as s biochemical sensing and surface enhanced Raman spectroscopy (SERS) [23-41], scanning near-field optical microscopy (SNOM) [4-8, 42, 43], near-field nano-lithography [44-46], novel light sources [2, 47-49], nonlinear optics [50-53], particle trapping [54-56], detection [57, 58], meta-materials [59], and many more.

The underlying physics, which is the heart of many of these applications, is the interaction of light with metallic nanostructures and the excitation of surface plasmon-polaritons. Throughout the last decade, the interaction of light with a wide variety of metallic nanostructures, such as nano-particles and shells [60], ellipsoids, triangular structures, dipoles, bowtie nano-antennas, crosses [61], nano-rings, split-ring resonators and many more (for a complete review see ref. [62] and references therein), has been thoroughly studied. More complex metallic structures such as plasmonic lenses [63-67], plasmonic mirrors [68, 69], nano-hole arrays [70-72], and tapered tips [73-76], have been proposed and studied as well for attaining nano-scale light focusing.

It is interesting to note that although the impact of the geometry of the nanostructures and the material composing them on the field enhancement was studied thoroughly, relatively little efforts were devoted to the impact of the properties of the impinging light. In most studies, the excitation field was either a plane wave or a focused Gaussian beam which were either linearly or circularly polarized. Recently, it was shown that radially polarized light can be coupled more efficiently to plasmonic lenses and produce brighter "hot-spot" than those produced by linearly polarized light [65, 77].

Radially polarized beams belong to an interesting family of beams known as circular vector beams (CVBs). CVBs [78] are unique beams which exhibit spatially dependent polarization state. Such beams exhibit linear polarization at each point across the beam profile but with varying orientation according to the beam type [78]. CVBs are formal solution of the vector wave equation in cylindrical coordinates, and are often categorized as either radially or azimuthally polarized.

CVBs have been the focus of intense studies due to their unique properties and diverse applications. In addition to efficient excitation of plasmons [67], radially polarized light is highly useful for Plasmonic imaging [79, 80], focusing beyond the diffraction limit [81-84] and optical trapping [85]. There are several methods for producing CVBs such as birefringent masks converting linear polarization beams into CVBs [86], diffractive optical interferometers [87], sub wavelength dielectric [88] and metallic [89] gratings, etc. More recently it was shown that radial Bragg lasers can spontaneously emit such beam [90].

In this paper, we study theoretically, using finite difference time domain (FDTD) simulations, the enhancement of the field in the gap of Au split-ring resonators (SRRs – see Fig. 1(a)) illuminated by *azimuthally polarized* beams, i.e. beams in which the electric field at each point is oriented orthogonally to the radius vector from the beam axis (see Fig. 1(b)). We find that when illuminated by azimuthally polarized beam, the intensity enhancement in the gaps of the SRR can be increased by more than an order of magnitude compared to that attained by linearly polarized beam. In addition, we find that the resonance of the SRR under azimuthal polarization is red-shifted (see also section 2) and that its bandwidth is narrower compared to that o identical structure under linear polarization illumination. Note, however that the Q-factor of the SRR *does not* depend on the polarization of the excitation and, therefore, cannot explain the dramatic increase in the intensity enhancement.

Figure 1(a) depicts a schematic of the SRR structure studied here as well as the definition of the geometrical parameters. SRR based structures have been studied by several groups and have been proposed and demonstrated for various applications. SRR [91-106] have been studied extensively as building blocks for metamaterials, sensing, transmission enhancement through metal films, slow light and, in particular, for attaining sharp spectral response similar to the phenomenon of electromagnetically induced transparency [107-114].

The SRR studied here consist of an Au ring, positioned on a $SiO_2$ substrate, from which two identical sections positioned on the opposite sides of a ring diameter are removed (see Fig. 1(a)). The structure is illuminated from the top by a Gaussian shaped beam which is either linearly or azimuthally polarized. The intensity of the field at the gap center, as a function of the frequency, is compared to its level without the SRR in order to evaluate the intensity enhancement. The rest of the paper is organized as follows: in section 2 we study the field enhancement and spectral properties of SRR illuminated by azimuthally polarized beams and compare them to those of linearly polarized beam illuminated structures. In section 3 we discuss the impact of the design parameters and in section 4 we present a method for tuning the SRR resonance frequency without changing its size by introducing additional gaps. In section 5 we study the impact of structure-beam registration errors on the SRR response and in section 6 we summarize the results and conclude.

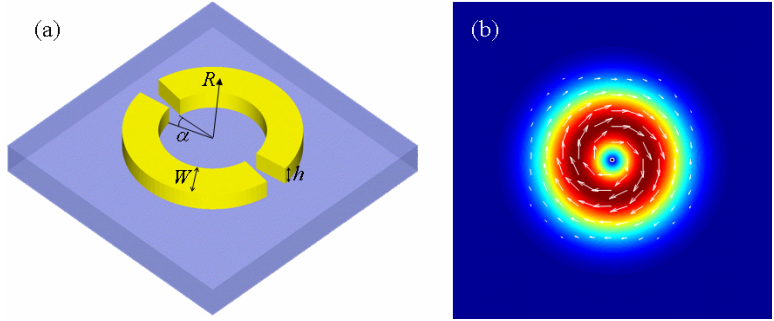

Fig. 1. (a) The SRR structure and parameter definitions; (b) The intensity and field profiles of azimuthally polarized beam.

## 2. Field enhancement and spectral properties: Azimuthally vs. linearly polarized beam

Figure 1(a) depicts a schematic of the SRR structure as well as the geometrical parameters defining the structure: $R$ – the SRR mid radius, $W$ – the SRR width, $h$ – the Au layer thickness and $\alpha$, which is the angular gap (actual gap is given by $R\cdot\alpha$). Figure 1(b) depicts a representative image of the intensity and field distribution of an azimuthally polarized beam. At each point, the electric field is *linearly* polarized but its vector is pointing orthogonally to the radius vector from the beam axis. Thus, in contrast to linearly or circularly polarized beam, the polarization state is inhomogeneous.

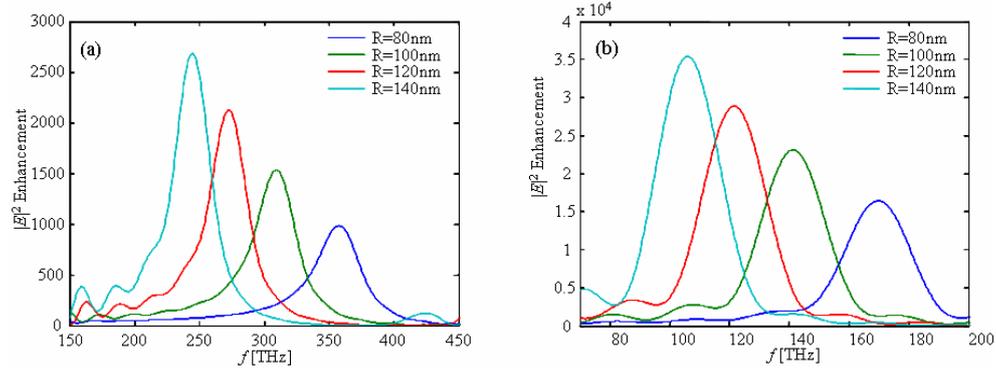

Fig. 2. Spectral intensity enhancement in SRRs with 14nm gaps for (a) linearly and (b) azimuthally polarized excitations.

The fundamental azimuthally polarized beam, which is the one pertaining to this paper, has the following form [115]:

$$E(\rho,z) = \hat{\phi} F(\rho,z) \exp[i(kz - \omega t)] \qquad (1)$$

where $k$ and $\omega$ are respectively the wavenumber and angular frequency and F is given by:

$$F(\rho,z) = A J_1\left(\frac{\beta\rho}{1+i\,z/L}\right) \cdot f(\rho,z) \cdot Q(\rho,z) \qquad (2)$$

where $A$ is a constant, $\beta$ is an arbitrary scale factor, $f$ is the fundamental Gaussian profile defined in (3), $L$ is the Rayleigh length given by $L = k\omega_0^2/2$ ($\omega_0$ is the waist) and $J_1$ is the order 1 Bessel function of the first kind. $Q$ is a phase term defined by Eq. (4).

$$f(\rho, z) = \frac{\omega_0}{\omega(z)} \exp[-i\Phi(z)] \exp\left(\frac{\rho^2/\omega_0^2}{1+i\,z/L}\right) \quad (3)$$

$$Q(z) = \exp\left(\frac{i\beta^2 z/2k}{1+i\,z/L}\right) \quad (4)$$

where $\omega^2(z)=\omega_0[1+(z/L)^2]$ is the beam width as a function of $z$ and $\Phi(z)=\tan^{-1}(z/L)$.

Fig. 2 depicts the spectral intensity enhancement, calculated by 3D FDTD simulation tool, in the gaps of SRR with various radii for linearly (Fig. 2(a)) and azimuthally (Fig 2(b)) polarized Gaussian beams. The SRR cross-section is 30nm×20nm and the mid-radius is varied between 80nm to 140nm. The angular gap is 10° for the 80nm radius SRR (corresponding to actual size of 14nm) where the actual gap is maintained constant for each SRR [i.e. the angular gap is $\alpha(R)=10\cdot 80/R$ degrees]. Fig. 3 depicts the intensity pattern (in logarithmic scale) in the 100nm radius SRR when excited on resonance with a linearly (Fig. 3(a)) and azimuthally (Fig. 3(b)) polarized light. As can be expected, in both cases the intensity is maximal in the gaps. However, for the linearly polarized excitation, there are also relatively strong fields in the sections of the SRR which are orthogonal to the gaps. This is because of currents which are excited along the narrow dimension of the ring for the $y$ polarized field. For the azimuthally polarized excitation, the intensity is concentrated in the gaps (and the vicinity) because the currents excited in each section of the SRR are in the same direction (azimuthal) and contribute to the build up of the intensity in the gaps.

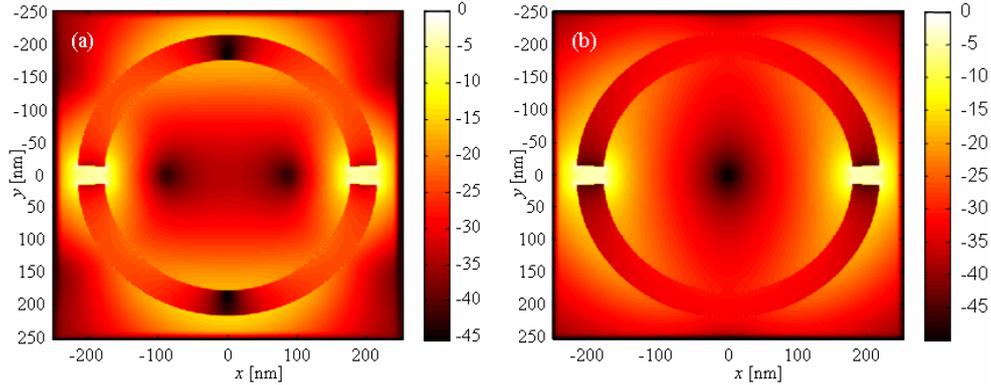

Fig. 3. Intensity pattern (log scale) in a 100nm radius SRR for (a) linear polarization excitation and (b) azimuthal polarization excitation.

Referring to Fig. 2, the comparison between the two excitation polarizations shows two eminent differences: 1) the intensity enhancement in the gap under azimuthally polarized light is larger by an order of magnitude than that for linearly polarized light; 2) the resonance wavelength of the SRR for azimuthally polarized beam is substantially longer than the resonance for linearly polarized light. The second property can be understood quite intuitively. Because of the curvature of the metallic SRR, the linearly polarized beam can efficiently excite currents only in parts of the metallic structure. Thus, for linearly polarized light the SRR acts as split-dipole antenna with an effective length which is determined primarily by the radius. For larger radii, the effective length of the dipole is longer which is manifested by a longer resonance wavelength – as can be seen in Fig. 2(a). On the other hand, azimuthally polarized light can efficiently excite currents throughout the whole SRR structure, thus effectively "seeing" a longer effective dipole length (approximately half the perimeter of the SRR).

Because of the SRR geometry, it seems reasonable that under azimuthally polarized illumination it can be considered roughly as an infinite 1D array of metallic nano-wires ($\pi-\alpha$)·$R$ long which are separated by $\alpha\cdot R$ air gaps. Figure 4 depicts the spectral intensity enhancement for nano-wire arrays which are "equivalent" to the SRRs of Fig. 2(b). The resonance frequencies of the SRRs and the "equivalent" arrays are almost identical where the small deviations can be easily understood by the fact that the approximation becomes less appropriate for smaller radii (note that the deviation is *smaller* for larger radii). Thus, it is clear that the red-shift of the resonance frequencies for azimuthal polarization stems from the effectively longer metallic structures.

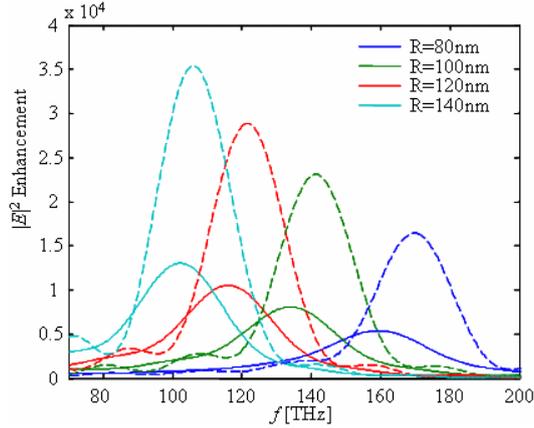

Fig. 4. Spectral intensity enhancement for nano-wire arrays which are "equivalent" to the SRRs of Fig. 2(b) (solid). The dashed lines spectra are identical to Fig. 2(b)

The second property (the substantially larger field enhancement) is less obvious. It has been shown that dense arrays of coupled metallic nano-resonators and antennas can modify the resonance wavelength and exhibit larger field enhancement than that of an isolated unit-cell [116, 117]. Although the "equivalent" arrays exhibit larger field enhancements than that of the isolated structure, this simplified model does not provide a complete insight to the much larger increase in the field enhancement observed in the azimuthal polarization illuminated SRR. We attribute this additional, unexpected, field enhancement to the better overlap between the spatial beam profile of the azimuthally polarized Gaussian beam (Fig. 1(b)) and the SRR structure. In particular, the formation of the null in the center of the beam (necessitated by the continuity of the fields) pushes the maximum of the field from the beam center towards the SRR metallic nano-wires, thus allowing more efficient collection and focusing of the beam power.

## 3. Impact of the SRR geometry

In addition to the SRR radius, the important design parameters are the ring width and the size of the gap. Intuitively, based on the known results of simple dipoles it can be expected that smaller gaps will produce larger field enhancements. The impact of the width is somewhat more complex but relying on the common knowledge that smaller features tend to produce larger enhancements one may expect to obtain stronger fields from narrower SRR.

Figure 5 depicts the resonance frequencies and intensity enhancements for SRRs of various radii with 8.4nm gaps (corresponding to $\alpha=6°$ for the 80nm radius SSR). Compared to the 14nm gaps SRRs of Fig. 2, the smaller gap induces larger field enhancement (by approximately a factor of 4 for the azimuthally polarized beams and only by a factor of 2.5 for linearly polarized beam) and a resonance shift of 7-10% for the azimuthal beam and 4-5% for the linear beam. Note, that the increase in the intensity enhancement due to the narrower gap is substantially larger for the SRR illuminated by azimuthally polarized beam.

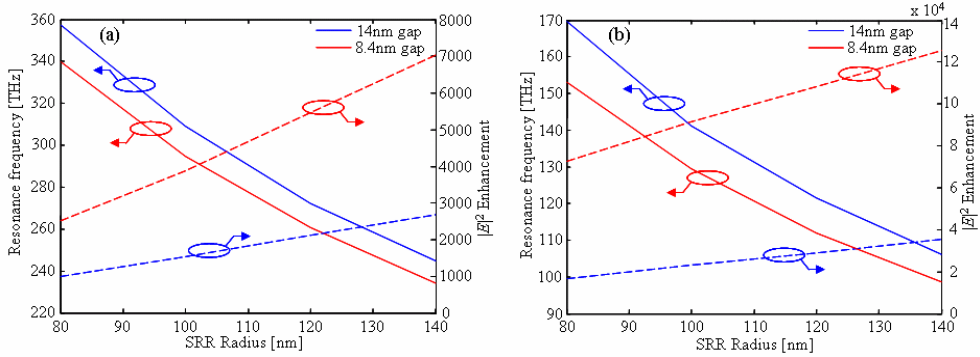

Fig. 5. Resonance frequencies and intensity enhancement in SRRs with 8.4nm gaps for (a) linearly and (b) azimuthally polarized excitations.

Figure 6 depicts the spectral responses and intensity enhancements for SRRs of various radii with 14nm gaps where the width of the ring was increased to 40nm. In contrast to the impact of the gap size, here there is a clear distinction between the azimuthal and the linear excitations. Compared to the response of the structure depicted in Fig. 2, the wider SRRs exhibit a clear blue-shift of their resonance for both polarization excitation types (approximately 10%). On the other hand, the impact of the SRR width on the intensity enhancement is completely different for the two types of the excitations. While the linearly excited SRRs exhibit larger intensity enhancement for the narrower structures (by approximately a factor of 2.3), the azimuthally excited structure exhibit larger enhancement for the *wider* SRR (by a factor of 2). This is a rather surprising result which is completely opposite to the intuitive expectation and the common knowledge. Similarly to the unexpected field enhancement, we attribute it to the better overlap between the spatial beam profile of the azimuthally polarized Gaussian beam and the SRR structure.

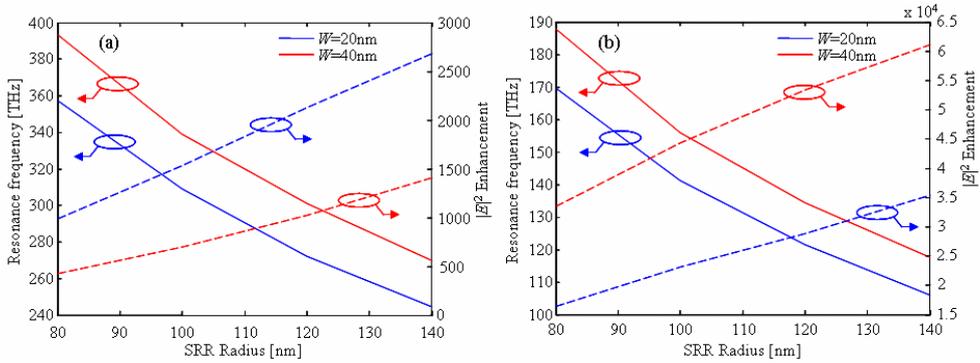

Fig. 6. Resonance frequencies and intensity enhancement in 20nm and 40nm wide SRRs with 14nm gaps for (a) linearly and (b) azimuthally polarized excitations.

## 4. Tuning the SRR resonance wavelength

Although the azimuthal polarization illumination provides a substantially lager field enhancement, it also induces a dramatic red-shift in the SRR resonance frequency. This is because of the longer effective length of the structure which interacts with the radiation. In order to re-tune the resonance frequency back to shorter wavelengths (which is desired in many cases) the length of the interacting metal structure should be reduced. However, in order to shift the resonance from, e.g. 130THz (for the 100nm SRR) towards 300THz the radius of the SRR should be decreased to approximately 30nm, thus making it practically extremely difficult.

Alternatively, it is possible to modify the resonance frequency without changing the SRR radius by introducing additional gaps to the structure (inset of Fig. 7) [118]. Figure 7(a) depicts the spectral intensity enhancement for a 100nm radius SRR with 4 gaps of 6nm. Note, that the resonance frequency shifts to 260THz due to the shorter arms of the nano-wires comprising the SRR. The intensity enhancement, however, is lower by approximately a factor of 2 as can be expected by the fact that the number of gaps (or focal energy points) is doubled. Fig. 7(b) shows the spectral enhancement for a linear polarization excitation exhibiting negligible shift in the resonance frequency.

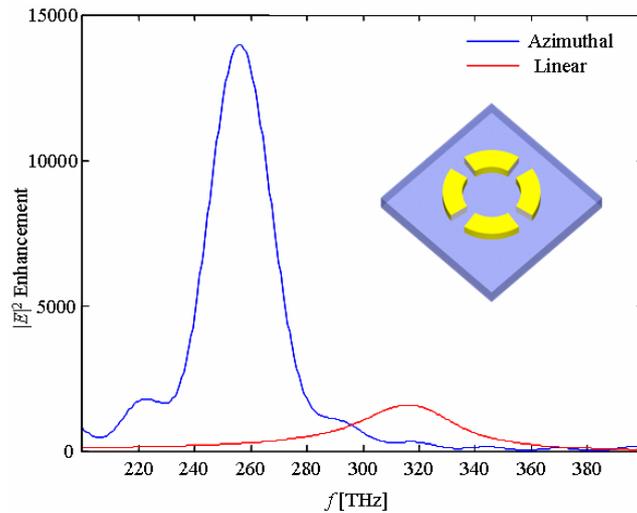

Fig. 7. Spectral intensity enhancement for a 100nm radius SRR with 4 gaps for azimuthal (blue) and linear (red) polarization. Inset – SRR geometry

Thus, by modifying the number of gaps in the SRR and their sizes, it is possible to tune the resonance frequency of the structure (for azimuthal illumination) without necessitating any changes in the SRR radius or in any of the critical dimensions of the structure.

### 5. The impact of the beam-structure registration errors

It is clear that in order to attain the maximal field enhancement from SRR illuminated by azimuthally polarized beam, it is desired that the beam center overlaps with that of the SRR. However, at nanometer scales such registration is quite difficult and it is expected that in practice some deviations might occur. The objective of this section is to study the impact of such deviations on the field enhancement in the SRR gaps.

Figure 8 depicts intensity enhancement in the gaps of a 100nm SRR with 4 gaps as a function of the difference between the beam and SRR centers. As shown in the figure, the enhancement drops when the beam and the SRR are not perfectly aligned. Nevertheless, the reduction in the enhancement is not dramatic – for a difference of 50nm (which is half the radius of the SRR) the intensity enhancement drop by merely 26%. We attribute this insensitivity to the beam-SRR registration to the fact that at resonance, the effective collection cross-section of the SRR is substantially larger than its physical dimensions, thus enabling the concentration of most of the optical power.

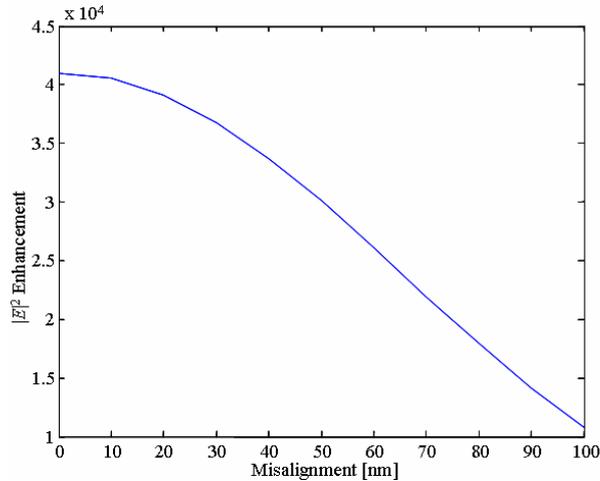
Fig. 8. The dependence of the intensity enhancement in the gap of a 100nm radius SRR on the beam-SRR centers registration error

### 6. Conclusions

We study the response of metallic split-ring resonator illuminated by azimuthally polarized beam. Compared to linearly polarized illumination, the intensity enhancement under azimuthal polarization is larger by more than an order of magnitude. We attribute this improvement in the intensity enhancement to two mechanisms: 1. The improved overlap between the electric field vector and the metallic resonator structure; 2. The improved overlap between beam profile the structure with enhances the effective absorption cross-section of the SRR.

The first mechanism increases the intensity enhancement by taking advantage of the complete perimeter of the SRR. Linearly polarized light induces current only in the SRR sections which are approximately parallel to the electric field. On the other hand, the azimuthally polarized light excites currents along the whole SRR which in turn result in a larger field in the gaps. The second mechanism increases the enhancement because of the intensity profile of the azimuthally polarized beam, which has an annular shape. Consequently, the SRR structure is capable of absorbing and focusing more power from the azimuthally polarized Gaussian beam which, again, results in larger field in the gaps.

Additionally, we studied the impact of the SRR geometrical parameters, specifically the size of the gaps and the width of the SRR, on the intensity enhancement. Interestingly, although smaller gaps result in large intensity enhancement (as can be expected), the increase in the enhancement is substantially larger when the SRR is illuminated by an azimuthally polarized beam. Another surprising result is the impact of the SRR width. While for structures illuminated by linear polarized beam, wider SRR results in *smaller* intensity enhancement, this trend is *reversed* when azimuthal polarization is employed.

We also present and study a simple approach to tune the resonance frequency of the SRR under azimuthally polarized illumination by introducing more gaps into the structure. The additional gaps basically decrease the effective length of the nano-wire sections comprising the SRR and consequently reduce the resonance wavelength. The intensity enhancement in each gap, however, decreases approximately linearly with the number of gaps, simply because the power is focused into more points. Nevertheless, it is possible to circumvent this drawback by designing an SRR with non-identical sections such as an SRR with three gaps, having two identical sections and a longer/shorter one. Such SRR is expected to exhibit three resonant modes with unequal intensities in the gaps. In particular, because of symmetry issues, one of these modes is expected to exhibit high intensity in one of the gaps and relatively low intensity in the other two. However, a complete study of such asymmetric structures is beyond the scope of this paper.

Practically, a perfect alignment between the beam and the SRR is difficult to attain. Therefore, we studied the sensitivity of the intensity enhancement in the gaps to misalignment between the axes of the azimuthally polarized beam and the SRR. We found that even for relatively large misalignments, the enhancement of the intensity does not drop rapidly and no change is observed in the resonance frequency and the bandwidth. Therefore, large intensity enhancements can still be attained even if the beam and the SRR are not perfectly aligned, thus rendering the proposed scheme practical and useful. The ability to substantially increase the field enhancement simply by tailoring the polarization property of the beam to the geometry of the structure is highly attractive for several applications such as SERS and NSOM.

**Acknowledgment**
The authors thank the Israeli Department of Defense and the Ministry of Science and Technology for partially supporting this research.